\documentclass[prl,twocolumn,showpacs,preprintnumbers]{revtex4}
\usepackage{graphicx}
\begin{document}

\title{Bound spinons in an antiferromagnetic ${\bf S}$=$\frac{1}{2}$ chain with a staggered field}

\author{M. Kenzelmann$^{1,2}$, Y. Chen$^{1,3}$, C. Broholm$^{1,2}$,
D.~H. Reich,$^{1}$ and Y. Qiu$^{2,4}$}

\affiliation{(1) Department of Physics and Astronomy, Johns
Hopkins University, Baltimore, Maryland 21218\\(2) NIST Center for
Neutron Research, National Institute of Standards and Technology,
Gaithersburg, Maryland 20899\\
(3)Los Alamos National Laboratory, Los Alamos, NM 87545\\
(4)Department of Materials Science and Engineering, University of Maryland, College Park, Maryland, 20742.}
\date{\today}

\begin{abstract}
Inelastic neutron scattering was used to measure the magnetic
field dependence of spin excitations in the antiferromagnetic
$S$=$\frac{1}{2}$ chain ${\rm CuCl_2 \cdot 2(dimethylsulfoxide)}$
(CDC) in the presence of uniform and staggered fields. Dispersive
bound states emerge from a zero-field two-spinon continuum with
different finite energy minima at wave numbers $q$=$\pi$ and
$q_i\approx\pi(1-2\langle S_z \rangle)$. The ratios of the field
dependent excitation energies are in excellent agreement with
predictions for breather and soliton solutions to the quantum
sine-Gordon model, the proposed low-energy theory for
$S$=$\frac{1}{2}$ chains in a staggered field. The data are also
consistent with the predicted soliton and $n$=$1,2$ breather
polarizations and scattering cross sections.
\end{abstract}
 \maketitle

Shortly after the advent of quantum mechanics, Hans Bethe
introduced a model antiferromagnet that continues to play a
central role in quantum many body physics \cite{bethe}. The
isotropic antiferromagnetic (AF) spin-1/2 chain, has a simple spin
Hamiltonian: ${\cal H}=J\sum_n{\bf S}_n\cdot{\bf S}_{n+1}$, is
integrable through Bethe's Ansatz, and is realized with high
fidelity in a number of magnetically anisotropic Cu$^{2+}$ based
materials. Because it sits at the boundary between quantum order
and spin order at $T=0$, Bethe's model is ideally suited for
exploring quantum critical phenomena and the qualitatively
different phases that border the critical point
\cite{sachdev,chainreview}. This letter presents an experimental
study of the profound effects of a symmetry breaking staggered
field on excitations in the spin-1/2 chain.

In zero field, the fundamental excitations of the spin-1/2 chain
are not spin waves but domain wall like quasi-particles called
spinons that separate reversed AF domains
\cite{Faddeev_Takhtajan,Muller,Karbach_Bougourzi}. The ground
state is a  Luttinger liquid and the spinons are non-local
spin-1/2 objects with short range interactions. Thus, spinons can
only be excited in pairs and produce a gapless continuum. Such a
spectrum has been observed in several quasi-one-dimensional
spin-1/2 chain systems \cite{Tennant,DenderPRB,Stone} and is now
understood to be a distinguishing attribute of quantum-critical
systems. The dramatic effect of a staggered field was discovered
through a high field neutron scattering experiment on the
quasi-one-dimensional spin-1/2 antiferromagnet copper benzoate
\cite{DenderPRL}. Designed to verify theoretical predictions of a
field driven gapless incommensurate mode \cite{Muller}, this
experiment instead revealed a field induced gap in the excitation
spectrum. The critical exponent of $\sim 2/3$ describing the field
dependence of the gap in copper benzoate, $\rm
[PM·Cu(NO_3)_2\cdot(H_2O)_2]_n$ \cite{Feyerherm}, and $\rm
Yb_4As_3$ \cite{Kohgi}, identified the source of this gap as the
staggered field that accompanies a uniform field in materials with
alternating Cu-coordination. The staggered field yields an
energetic distinction between reversed domains, which confines
spinons in multi-particle bound states \cite{Oshikawa_Affleck}.

A quantitative theory for this effect was developed by Affleck and
Oshikawa starting from the following extension of Bethe's model
Hamiltonian \cite{Oshikawa_Affleck,Affleck_Oshikawa},
\begin{eqnarray}
\mathcal{H} = J\sum_i {\bf S}_i\, {\bf S}_{i+1}
+\sum_j (-1)^j {\bf D}\cdot ({\bf S}_{j-1}\times{\bf S}_j)&\nonumber \\
- \sum_{j,\alpha,\beta} H^\alpha [g^u_{\alpha\beta} + (-1)^j
g^s_{\alpha\beta}]S^\beta_j  \, . \label{Hamiltonian}
\end{eqnarray}
The alternating spin environment is represented by the staggered
Dzyaloshinskii-Moriya (DM) interaction and Zeeman terms. Through
an alternating coordinate transformation, the model can be mapped
to a spin-1/2 chain in a transverse staggered field that is
proportional to the uniform field $H$. While the zero field
properties of Eq.(1) are indistinguishable from Bethe's model, an
applied field induces transverse AF Ising spin order and a gap.
Using bosonization techniques to represent the low energy spin
degrees of freedom, Affleck and Oshikawa showed that their
dynamics is governed by the quantum sine-Gordon model (QSG) with
Lagrangian density
\begin{eqnarray}
{\cal L}=\frac{1}{2}[(\partial_t\overline{\phi})^2-(\partial_x\overline{\phi})^2 ]+hC\cos(\beta\overline{\phi}).
\end{eqnarray}
Here $h\propto H$ is the effective staggered field. $C(H)$ and
$\beta(H)$ (which goes to $ \sqrt{2\pi}$ for $H\rightarrow 0$)
vary smoothly with the applied field and can be determined
numerically through the Bethe Ansatz for $h<<H$
\cite{Affleck_Oshikawa,Essler98}.

With applications from classical to particle physics, the SG model
plays an important role in the theory of non-linear dynamic
systems \cite{Tsvelik_book}. Excitations are composed of
topological objects called solitons that encompass a localized
$\pm 2\pi/\beta$ shift in $\overline{\phi}$ for a soliton and
anti-soliton respectively \cite{Dashen_Hasslacher}. In addition
there are soliton-antisoliton bound states called breathers, which
drop below the soliton-anti-soliton continuum as the  non-linear
term is increased. The excited state wave functions are known for
both solitons and breathers, which enables exact calculation of
the inelastic scattering cross sections. In this Letter we use
neutron scattering from a magnetized spin-1/2 chain with two spins
per unit cell to test these results and more generally to explore
the dynamics of spinons with long range interactions.

Based on the temperature dependence of the susceptibility and
specific heat, ${\rm CuCl_2 \cdot 2((CD_3)_2SO)}$ (CDC) was
identified as an AF $S$=$\frac{1}{2}$ chain system with
$J$=$1.5\;\mathrm{meV}$, a staggered $g$-tensor and/or DM
interactions \cite{Landee,Chen_CDC}. The spin chains run along the
${\bf a}$-axis of the orthorhombic crystal structure ({\it Pnma})
\cite{Willett_Chang}, with the ${\rm Cu^{2+}}$ ions separated by
$0.5{\bf a} \pm 0.22{\bf c}$. Wave vector transfer is indexed in
the corresponding reciprocal lattice ${\bf Q}(hkl)=h{\bf
a}^*+k{\bf b}^*+l{\bf c}^*$, and we define $q={\bf Q}\cdot {\bf
a}$. Due to weak inter-chain interactions, CDC has long-range AF
order in zero field below $T_N$=$0.93\;\mathrm{K}$ with an AF
wave-vector ${\bf Q}_m={\bf a}^*$. An applied field strongly
suppresses the ordered phase \cite{Chen_CDC}, indicating that
inter-chain interactions favor correlations that are incompatible
with the field-induced staggered magnetization\cite{Oshikawanew}.
Above the $H_c = 3.9\;\mathrm{T}$ critical field for N\'{e}el
order, we find that CDC is an excellent model system for our
purpose.

\begin{figure}[t]
\begin{center}
  \includegraphics[height=9.5cm,bbllx=38,bblly=137,bburx=527,
  bbury=688,angle=0,clip=]{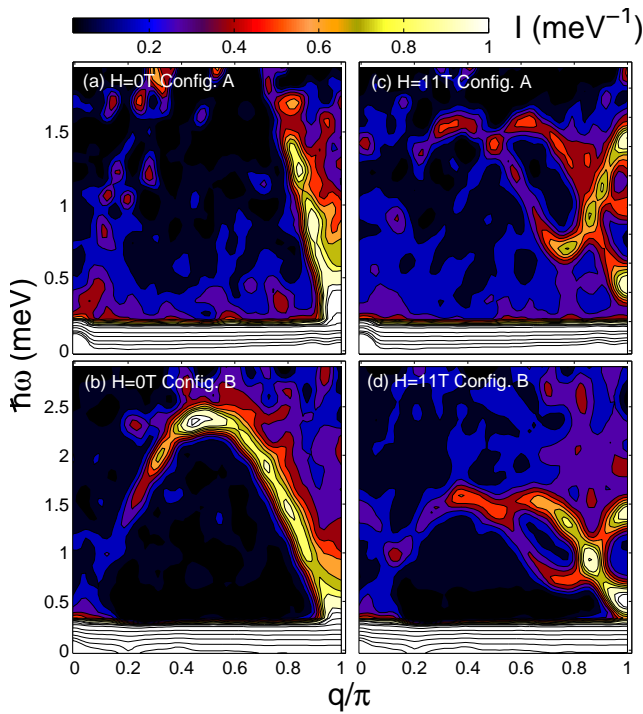}
  \caption{Normalized scattering intensity $I(q,\omega)$
  in zero field and $H$=$11\;\mathrm{T}$ in the $S$=$\frac{1}{2}$
  chain CDC as a function of the chain wave-vector transfer $q$
  and energy transfer $\hbar\omega$, measured at $T$=$40\;\mathrm{mK}$.
  The data were binned and smoothed, leading to an effective
  wave-vector resolution $\delta q$=$0.057\pi$ and
  $0.061\pi$ and an energy resolution $0.1$ and $0.18\;\mathrm{meV}$
  for configurations A and B, respectively. The colorbar indicates
  the scattering strength.}
  \label{Fig1-colorplot}
\end{center}
\vspace{-0.25in}
\end{figure}

Deuterated single crystals were grown through slow cooling of
saturated methanol solutions of anhydrous copper chloride and
deuterated dimethyl sulfoxide ($\rm (CD_3)_2SO$) in a 1:2 molar
ratio \cite{Chen_CDC}. The sample studied consisted of four
crystals with a total mass $7.76\;\mathrm{g}$. The experiments
were performed using the disk chopper time-of-flight spectrometer
(DCS) at the NIST Center for Neutron Research with the ${\bf
c}$-axis and the magnetic field vertical. In configuration A the
incident energy was $E_i$=$3.03\;\mathrm{meV}$ and the ${\bf
a}$-axis was parallel to the incident neutron beam direction ${\bf
k}_i$. Configuration B had $E_i$=$4.64\;\mathrm{meV}$ and
$\angle({\bf k_i},{\bf a})$=$60^{o}$. The counting time was 18 hrs
at 11 T and an average of 5 hrs for each measurement between 0 and
8T. The raw scattering data were corrected for a time-independent
background measured at negative energy transfer, for monitor
efficiency, and for the ${\rm Cu^{2+}}$ magnetic form factor,
folded into the first Brillouin zone, and put onto an absolute
scale using the elastic incoherent scattering from CDC. For the
normalization, the H/D ratio (=0.02) was measured independently
through prompt-$\gamma$ neutron activation analysis.

\begin{figure}[t]
\begin{center}
  \includegraphics[height=7.7cm,bbllx=41,bblly=151,bburx=545,
  bbury=696,angle=0,clip=]{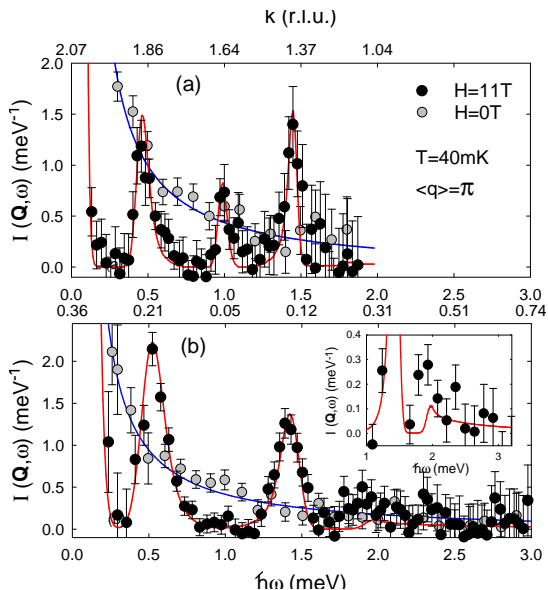}
  \caption{Spectra at $q$=$\pi$ observed with configurations A
  (a) and B (b) for zero field and $11\;\mathrm{T}$ obtained
  from data in the range $0.95 \pi < q < 1.05\pi$. The zero-field
  intensity for  $0.3\pi<q<0.4\pi$ was subtracted as a background.
  The top axes indicates wave-vector transfer, $k$, perpendicular
  to the chain. The blue line is a fit of the two-spinon cross-section
  to the zero field scattering \protect\cite{Karbach_Bougourzi}.
  The red lines are fits to the $H=11$ T data as explained in the text,
  including the theoretically calculated breather continua
  polarized perpendicular to the field \protect\cite{Essler98}.
  The inset in (b) highlights the high energy range. }
  \label{Fig2-piscans}
\end{center}
\vspace{-0.25in}
\end{figure}

\begin{figure}[t]
\begin{center}
  \includegraphics[height=7.2cm,bbllx=85,bblly=230,bburx=510,
  bbury=625,angle=0,clip=]{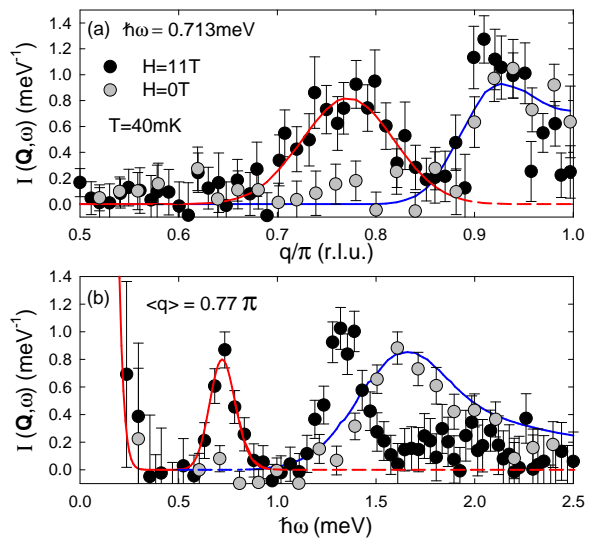}
  \caption{Constant-$\omega$ (a) and constant-$q$ scans (b) through
  the gapped soliton at $H$=$11\;\mathrm{T}$ at the incommensurate
  wave-vector $q_i$. The zero-field scattering is shown for
  comparison together with a fit of the two-spinon cross-section
  to the data. The unfitted peak corresponds to a moving breather.
  These data are from configuration B. The ranges of integration
  were (a) 0.675 meV $< \hbar \omega < $  0.75 meV, and  (b)
  $0.72 \pi < q < 0.82 \pi$. }
  \label{Fig3-incompeak}
\end{center}
\vspace{-0.25in}
\end{figure}

Figures \ref{Fig1-colorplot}(a) and \ref{Fig1-colorplot}(b) show
that for $T \ll J/k_{\rm B}$, the zero-field excitation spectrum
of CDC consists of continuum scattering above a low-energy
threshold that varies as $\hbar\omega$=$\frac{\pi}{2}J |\sin(q)|$
through the zone \cite{Muller}. An exact analytical expression for
the two spinon contribution to the scattering cross section which
accounts for 72.89\% of the total spectral weight was recently
obtained by Bougourzi {\em et al.}
\cite{bougourzi1996,fledderjohann1996,Karbach_Bougourzi}. Figures
\ref{Fig2-piscans} and \ref{Fig3-incompeak} show a quantitative
comparison of this result (blue line), duly convoluted with the
experimental resolution, to the experimental data. The excellent
quantitative agreement between model and data provides compelling
evidence for spinons in the zero field state of CDC. Note that the
Goldstone modes that are expected due to N\'{e}el order for
$\hbar\omega < k_{\rm B}T_N\approx 0.1\;\mathrm{meV}$, are not
resolved in this experiment.\par

\begin{figure}[t]
\begin{center}
  \includegraphics[height=5.1cm,bbllx=55,bblly=216,bburx=560,
  bbury=519,angle=0,clip=]{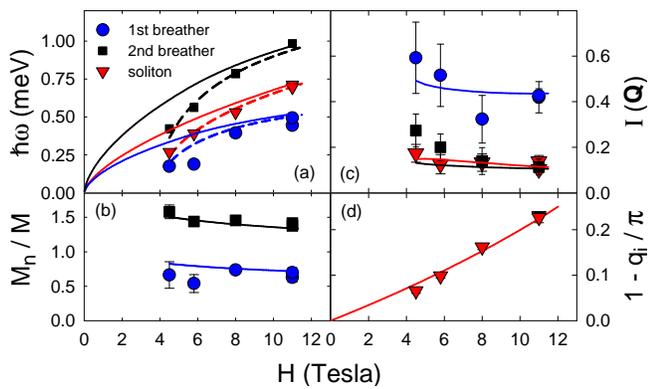}
  \caption{(a) Field dependence of the soliton and breather modes.
  The lines are the QSG predictions for
  $h\propto H$ (solid) and $h$=$c(H-H_c)^{0.68}$
  (dashed). (b) Ratio of breather
  energies to soliton energies versus $H$,
  compared to the QSG model. (c) Energy integrated
  intensities divided by the QSG predicted polarization factors and
  compared to structure factor predictions using particle energies
  given by the dashed line in (a) and a common overall scale
  factor \protect\cite{Essler98}. (d) Incommensuration $q_i$ obtained
  from the minimum in the soliton dispersion compared to the predicted
  $q_i=\pi(1-2\langle S_z \rangle)$ using the magnetization curve
  $\langle S_z \rangle=1/\pi\arcsin((1-\pi/2+\pi J/(g_z H \mu_B))^{-1})$
  for a Heisenberg spin-1/2 chain\cite{Muller}. The field dependence
  was measured with configuration A, and an additional measurement at
  $H=11\;\mathrm{T}$ was performed using configuration B.}
  \label{Fig4-fitresults}
\end{center}
\vspace{-0.25in}
\end{figure}

Figures \ref{Fig1-colorplot}(c) and \ref{Fig1-colorplot}(d) show
that the magnetic excitations in CDC change dramatically with
field and are dominated by resolution-limited modes for $H=11$ T.
Figures \ref{Fig2-piscans} and \ref{Fig3-incompeak}(b) show
spectra at the wave vectors corresponding to the minima in the
dispersion relations, which occur at $q = \pi$, and $q_i =0.77
\pi$, as determined from the constant-$\hbar\omega$ cut in
Fig.~\ref{Fig3-incompeak}(a). These data graphically illustrate
the field-induced transfer of spectral weight from the two-spinon
continuum into single-particle excitations. A phenomenological
cross-section of long-lived dispersive excitations was fit to the
data near $q=\pi$ and $q_i$ to take into account the experimental
resolution and thereby accurately locate the excitation energies.
These fits are shown as red lines in Figs.~\ref{Fig2-piscans} and
\ref{Fig3-incompeak}, and the inferred parameters characterizing
the dispersion relations are displayed for a series of fields in
Fig.~\ref{Fig4-fitresults}.\par

We now examine whether our high field observations are consistent
with the QSG model for spin-1/2 chains in a staggered field
\cite{Affleck_Oshikawa}. First, the model predicts single soliton
excitations at $q_i$=$\pi(1-2\langle S_z \rangle)$. This  is
qualitatively consistent with the raw data in Figs.~1(c) and 1(d),
and with Fig.~4(d), which shows how $q_i$ moves across the zone
with $H$. Quantitative agreement is also apparent from the solid
line in Figure 4(d), which is the predicted field dependence as
calculated from the magnetization curve for a spin-1/2 chain
\cite{Muller}.\par

The soliton and antisoliton mass is related to the exchange
interaction of the original spin chain and the uniform and
staggered fields, $H$ and $h$, as follows
\cite{Dashen_Hasslacher,Affleck_Oshikawa}
\begin{eqnarray}
    &M\approx J (\frac{g \mu_B h}{J})^{(1+\xi)/2} ~\times&\nonumber \\
    &\{B(\frac{J}{g \mu_BH})^{(2\pi-\beta^2)/4\pi}
    (2-\frac{\beta^2}{\pi})^{1/4}\}^{-(1+\xi)/2}. &
    \label{Eq2}
\end{eqnarray}Here $\xi=\beta^2/(8\pi-\beta^2)\rightarrow 1/3$ for
$H\rightarrow 0$ and $B=0.422169$. Assuming $h\propto H$, the
soliton energy versus field is shown as a solid red line in
Fig.~\ref{Fig4-fitresults}(a). While Eq. (\ref{Eq2}) is at the
limit of validity for CDC at $H=11$ T, we attribute the
discrepancy with the mode energy at $q_i$ (red triangles) to
inter-chain interactions that suppress the effective staggered
field close to $H_c$. The more general expression, $h \propto
(H-H_c)^{\alpha}$, yields a good fit for $\alpha=0.68(5)$ (dashed
red line).\par

The QSG model predicts that breather bound states of $2n$-solitons
should be accessible at $q$=$\pi$ with masses
\begin{equation}
    M_n=2M\sin(n\pi\xi/2)\, .
\end{equation}
Sharp modes are indeed observed in CDC at $q=\pi$.  Their energies
are compared to the breather masses predicted for $h \propto H$
(solid lines) and $h \propto (H-H_c)^{0.68}$ (dashed lines) in
Fig.~\ref{Fig4-fitresults}(a). The ratios of the commensurate and
putative breather mode energies to the lowest energy
incommensurate and putative soliton mode energy,  shown in
Fig.~\ref{Fig4-fitresults}(b), are in excellent agreement with the
normalized breather energies  $M_n/M$ for $n$=$1$ and $n$=$2$.
This comparison, which is insensitive to the origin of the
staggered field, suggests that breathers indeed exist in CDC.

The evidence for breathers is strengthened as we examine the
polarization of the scattering at $q=\pi$. According to the QSG
model, $n$=odd (even) breathers are polarized in the plane normal
to ${\bf H}$ and perpendicular (parallel) to ${\bf
h}$\cite{Essler98}. Neutron scattering probes the projection of
spin fluctuations on the plane normal to the scattering vector
${\bf Q}$. Figure \ref{Fig2-piscans} shows that the
$\hbar\omega$=$1\;\mathrm{meV}$ peak seen for ${\bf Q}_{\rm
A}\approx(1,1.64,0)$ in configuration A is absent for
$\hbar\omega$=$1\;\mathrm{meV}$ and ${\bf Q}_{\rm
B}\approx(1,0,0)$ in configuration B. In a quasi-one-dimensional
system the only explanation for this is that the excitation is
polarized along ${\bf Q}_{\rm B} ||{\bf a} || {\bf h}$ as expected
for an even numbered breather, and hence is extinguished by the
polarization factor in the neutron scattering cross section for
configuration B. The predicted ${\bf b}$ and ${\bf c}$ axis
polarizations respectively of the $n$=$1$ breather and the soliton
are confirmed by the consistent polarization factor corrected
intensities from configurations A and B for $H=11\;\mathrm{T}$ in
Fig.~\ref{Fig4-fitresults}(c).\par

One of the unusual aspects of the QSG model is that complex
features such as the breather and soliton structure factors can be
calculated exactly \cite{Essler98}. The solid lines in
Fig.~\ref{Fig4-fitresults}(c) show that these  exact results are
consistent with the field dependent intensities of the
commensurate and incommensurate low energy modes in CDC. For
$H=11$\ T the third breather is expected at about
$1.4\;\mathrm{meV}$, close to the energy of the soliton mode at
$h$=$1$. The peak close to $1.4\;\mathrm{meV}$ has intensity
$I=0.14(3)$ in configuration A and $I=0.26(2)$ in configuration B
and this is consistent with a third breather contribution
polarized along ${\bf b}$. The inferred ${\bf b}$-polarized
intensity of $I_b^{\rm exp}=0.23(7)$ is however much greater than
the intensity predicted for the $n=3$ breather ($I_3^{\rm
QSG}=0.026(7)$), which indicates additional anisotropic
contributions to the inelastic scattering there.

In addition to the field-induced resonant modes, the experiment
shows that a high energy continuum persists for
$H$=$11\;\mathrm{T}$. Fig.~\ref{Fig1-colorplot}(d) for example
clearly shows a broad maximum in the $q-$dependence of neutron
scattering for energies $\hbar\omega>1.6$ meV and $q\approx\pi$.
Firm evidence for continuum scattering comes from the field
dependence of the first moment $\langle\hbar\omega\rangle_{\bf
Q}$=$\hbar^2\int S({\bf Q},\omega)\omega d\omega$, which is
proportional to the ground state energy $\langle{\mathcal
H}\rangle$ with a negative $q$-dependent prefactor
\cite{Hohenberg_Brinkman}. At zero field the experimental value of
$\langle\hbar\omega\rangle_{\bf Q}$ corresponds to
$\langle{\mathcal H}\rangle=-0.4(1)J$, in agreement with Bethe's
result of $\langle{\mathcal H}\rangle=(\frac{1}{4}$-$\ln2)J\approx
-0.44J$. At $11\;\mathrm{T}$, however, $\tilde{\langle{\mathcal
H}\rangle}$ derived solely from the resonant modes is $-0.25(6)J$
when $\langle{\mathcal H}\rangle$ is expected to be $-0.34J$
\cite{Chen_CDC}. The discrepancy is an independent indication of
spectral weight beyond the resonant modes. For $q=\pi$, the
transverse contribution to the continuum scattering predicted by
the QSG model \cite{Essler98} and shown as a solid line in the
inset of Fig.~\ref{Fig2-piscans} has a maximum close to a weak
peak in the measured scattering intensity. The shortfall of the
theoretical result suggests that there are additional longitudinal
contributions to the continuum scattering.\par

In summary, staggered field induced spinon binding in spin-1/2
chains provides an experimental window on the unique non-linear
dynamics of the quantum sine-Gordon model. Our neutron scattering
experiment on quasi-one-dimensional CDC in a high magnetic field
yields clear evidence for soliton/antisoliton creation at wave
vector transfer $q=\pi(1-2\langle S^z\rangle)$, as well as $n=1$
and $n=2$ breather bound states at $q=\pi$. Interpretation of the
data throughout the Brillouin zone will require exact
diagonalization studies and a better understanding of lattice
effects than provided by the continuum field theory reviewed in
this paper. Other results that call for further experimental and
theoretical work are the observation of high energy continuum
scattering in the gapped phase and indications that inter-chain
interactions can renormalize the soliton mass.

\begin{acknowledgments}
We thank  C. P. Landee, J. Copley, C. Batista, I. Affleck, and  F.
Essler for helpful discussions and R. Paul for prompt gamma
analysis. Work at JHU was supported by the NSF through
DMR-0306940. DCS and the high-field magnet at NIST were supported
in part by the NSF through DMR-0086210 and DMR-9704257.
\end{acknowledgments}

\bibliographystyle{prsty}

\end{document}